# Highly Accurate Determination of Heterogeneously Stacked Van-der-Waals Materials by Optical Microspectroscopy

Andreas Hutzler[1,]*, Birk Fritsch[1], Christian D. Matthus[2], Michael P. M. Jank[3], Mathias Rommel[3]

[1]Electron Devices (LEB), Friedrich-Alexander University Erlangen-Nürnberg, Cauerstraße 6, 91058 Erlangen, Germany

[2]Circuit Design and Network Theory, Technische Universität Dresden, Helmholtzstraße 18, 01069 Dresden, Germany

[3]Fraunhofer Institute for Integrated Systems and Device Technology (IISB), Schottkystraße 10, 91058 Erlangen, Germany

*Corresponding author: andreas.hutzler@fau.de, ORCID: 0000-0001-5484-707X

**Abstract**

The composition of Van-der-Waals heterostructures is conclusively determined using a hybrid evaluation scheme of data acquired by optical microspectroscopy. This scheme deploys a parameter set comprising both change in reflectance and wavelength shift of distinct extreme values in reflectance spectra. Furthermore, the method is supported by an accurate analytical model describing reflectance of multilayer systems acquired by optical microspectroscopy. This approach allows uniquely for discrimination of 2D materials like graphene and hBN and, thus, quantitative analysis of Van-der-Waals heterostructures containing structurally very similar materials. The physical model features a transfer matrix method which allows for flexible, modular description of complex optical systems and may easily be extended to individual setups. It accounts for numerical apertures of applied objective lenses and a glass fiber which guides the light into the spectrometer by two individual weighting functions. The scheme is proven by highly accurate quantification of the number of layers of graphene and hBN in Van-der-Waals heterostructures. In this exemplary case, the fingerprint of graphene involves distinct deviations of reflectance accompanied by additional wavelength shifts of extreme values. In contrast to graphene the fingerprint of hBN reveals a negligible deviation in absolute reflectance causing this material being only detectable by spectral shifts of extreme values.

1.  **Introduction**

Optical spectroscopy is used for characterizing ultra-thin films down to atomic layers. It has been shown that in particular reflectance spectroscopy is well-suited for detecting single atomic layers of 2D materials like graphene (Gr)[1–6], hexagonal boron nitride (hBN)[3,7,8] or transition metal dichalcogenides (TMD)[3,8–10] like $MX_2$ (M = Mo, W; X = S, Se, Te). For this purpose, reflectance spectra of a dielectric layer stack in combination with a superimposed ultra-thin layer are acquired and the deviation between both reflectance spectra is evaluated in form of contrast spectra[5]. The optical contrast between 2D materials and subjacent (or overlying[11]) layer stacks can be used for determining the number of layers and is tailored by choosing appropriate combinations of materials and film thicknesses (i.e. tailoring optical pathways). This is achieved by predicting the reflectance behavior of the layer stack via analytical calculations while considering the optical properties of each individual layer[5,12]. Additionally, a spectral shift of distinct extreme values in reflectance spectra can be analyzed to gain information about the number of layers of a 2D material as well[8]. However, an appropriate discrimination between different stacked 2D materials in Van-der-Waals heterostructures remains challenging utilizing only one of both mentioned parameters (contrast or wavelength shift) because of the ambiguous origin of occurring changes. Furthermore, in order to obtain microscopic information with high spatial resolution, reflectance spectra are recorded by utilizing microscope setups deploying objective lenses with large magnification, often accompanied by large numerical apertures (*NA*). Such *NA*s dramatically influence measurement results due to the superposition of light reflected from a wide range of collection angles. Thus, another major challenge evolving during quantitative data analysis is to properly describe the optical setup in a modular way by flexible analytical models including the effective *NA*[13] as well as all other deteriorating optical elements between the sample and the spectrometer. In order to determine the composition of unknown Van-der-Waals heterostructures, the analytical model needs to provide a very high precision because exceptionally small changes in reflectance induced by single atomic layers of 2D materials have to be conclusively evaluated.

In this study, we present a hybrid approach for quantitative determination of the composition of Van-der-Waals heterostructures by evaluation of reflectance and spectral changes of microspectroscopic data. This quantitative analysis is achieved by enhancing and refining established analytical models[5,14] which is a prerequisite for utilizing them for determining small changes in both absolute reflectance and spectral position of distinct extreme values. For this purpose, the individual optical elements of the measurement setup are considered in the analytical model for calculating expected reflectance spectra of the optical layer stack under investigation. These optical elements include light source, objective lens (numerical aperture) and a glass fiber coupling the reflected light into the spectrometer. With the optical system being accurately described, different 2D materials in Van-der-Waals heterostructures can be discriminated while simultaneously determining their number of layers by combining both mentioned evaluation schemes, i.e. the contrast-based[5] (in this particular case absolute reflectance instead of contrast is deployed) and the wavelength-shift based method[8]. The applicability of this strategy is demonstrated for the extreme case of graphene-hBN heterostructures, where the influence of individual layers on reflected light is weak due to their extraordinary small thickness (graphene: 3.35 Å[2]; hBN: 3.33 Å[15]). In addition, hBN, which is also denoted as 'white graphene'[16], has a colorless appearance and is known to be faintly visible by optical methods. The presented approach is easily adaptable for arbitrary layer stacks and measurement setups employing magnifying objective lenses, glass fibers, polarized and/or incoherent light.

2. **Methods**

Analytical calculations are performed applying a generalized transfer matrix method[17] (TMM) which is implemented in Delphi™ utilizing complex refractive indices for calculating reflectance (and transmittance) spectra of multilayer systems as described elsewhere[5]. Note that the incidence angle as well as the polarization state of the light source is already included into this model. Complex refractive indices of air[18], silicon[19], graphene[20], hBN[21] and poly(methyl methacrylate) (PMMA)[22] were taken from literature and the complex refractive index of silicon dioxide was determined by spectroscopic ellipsometry (SE). The choice of proper refractive indices is crucial for achieving a high accuracy as needed in this extreme case. Offsets in *n* and *k* would result in a change in the wavelength shift and the absolute reflectance, respectively. The measurement setup used in this work (**Figure 1(a)**) is a Zeta 300 optical profiler from KLA Corporation, USA, which is equipped with an ultra-bright white LED light source, four different objective lenses (i.e. 20x magnification with *NA* = 0.4 and a point resolution of 12.5 µm, 50x with *NA* = 0.35 and 6.3 µm, 50x with *NA* = 0.8 and 6.3 µm and 100x with *NA* = 0.9 and 3.1 µm), a multimode glass fiber with a core diameter of 200 µm (*NA* = 0.07) which couples the reflected light onto a line spectrometer with a spectral measurement range of 410 nm to 790 nm and a spectral bandwidth Δλ of 2 nm that is operated in crossed Czerny-Turner configuration[23]. The light source is projected in a homogeneous profile onto the sample and has a bandwidth Δλ of 350 nm (wavelength λ of 400 nm to 750 nm) which corresponds to a nominal coherence length of about 940 nm for the center wavelength of 575 nm ($\lambda^2/\Delta\lambda$). This so-called Köhler illumination[24] describes the case of the central plateau of a point spread function (Airy disc[25]) caused by convolution of a dot-shaped light source with a circular aperture being projected onto the area of interest. Measurements were performed utilizing integration times of 2 s for acquiring reflectance spectra at two spots at least for each stack.

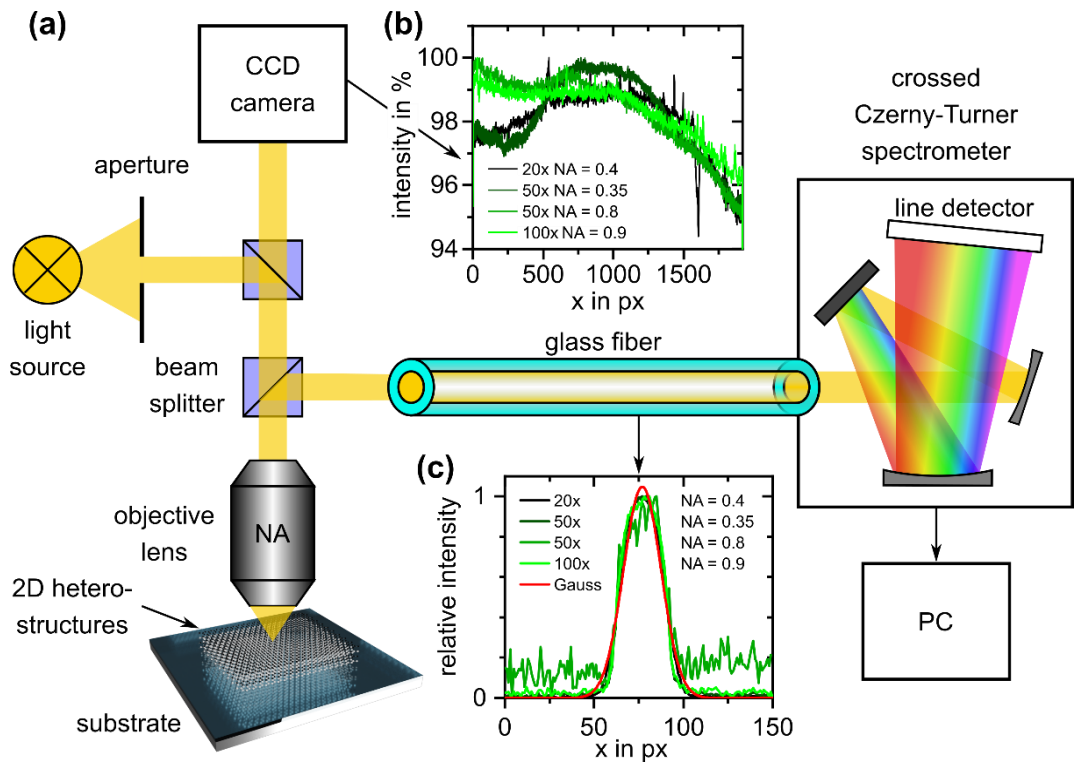

*Figure 1*. *(a) Schematics (simplified) of the measurement setup used in this work. Diagrams in (b) and (c) show illumination intensity distribution along the x-axis of the area of interest as well as transversal intensity distribution of light passing the glass fiber.*

## 3. Results and Discussion
### 3.1. Characterization and modeling of the utilized measurement system

An accurate evaluation of reflectance spectra recorded by a real measurement system requires pre-characterization of the optical setup. Thus, individual optical elements of the measurement setup were investigated and the results are taken into account by reasonable modeling. Although the modeling performed in this work is related to a particular setup, the general strategy is also applicable for optical modeling of different microspectroscopic systems. This is because the proposed analytical model which is based on a TMM, allows for modular description of individual elements and, thus, provides flexible modelling of the optical system. In order to determine the homogeneity of the illumination intensity, micrographs of a highly uniform surface with respect to spectral reflectivity were recorded through different objective lenses. The illumination homogeneity was found to be above 95% along 1920 pixels (**Figure 1(b)**). The observed small inhomogeneities, however, are neglected in further calculations because the area of the measurement spot from which the reflected light is coupled into the spectrometer is far smaller (only about 50 pixels in diameter which corresponds to approx. 0.3% of the field of view (FOV)) than the illuminated area which is projected onto the CCD camera of the microscope. The coherence length of the light source is considered in terms of incoherent illumination following the approach of Santbergen *et al.*[14] and can be estimated to be in the range of 940 nm as stated above. Nevertheless, measurements revealed no damping of interference patterns induced by partial incoherence of the light source in reflectance spectra of silicon oxide layers with a physical thickness of up to 1.5 µm. This is caused by the light dispersion in the spectrometer which reduces the light bandwidth to 2 nm per individual channel of the detector resulting in a much larger coherence length for each channel (e.g., 165 µm for a wavelength of 575 nm). In the presented case, the incoherence is, thus, neglected. Furthermore, the light source is assumed to emit non-polarized (i.e. 50% transverse electric (TE) mode and 50% transverse magnetic (TM) mode) light.

A major impact on measured reflectance spectra is found to be caused by objective lenses which are located in the ray path. This is caused by numerical apertures allowing for light acquisition also at larger angles of acceptance (i.e. $\alpha_{max} = \arcsin NA$) than in the case of normal incidence. In a first approximation, all different ray paths superimpose to an averaged spectrum. In order to take this effect into account, reflectance spectra are modelled by implementation of a weighting function suggested by Saigal *et al.*[13].

After reflection from the specimen and passing the objective lens again, the light is coupled into the spectrometer via a multi-mode glass fiber with a length of 50 cm. This corresponds to a convolution of the reflected light with a circular aperture (core of the glass fiber) which would result in a light beam with an intensity profile having the form of an Airy disc[25]. However, as the envelope of the glass fiber is not completely opaque and transmits the evanescent field of light located in the core, side maxima of the Airy disc are damped which leads to a beam profile that can be appropriately described by a Gaussian function[26]. This theoretical deliberation was practically validated by coupling light into the glass fiber from the spectrometer side for projecting the intensity distribution of light which passes the glass fiber through the objective lens onto the specimen. The results are presented in **Figure 1(c)**. Here, it can be seen that the beam profile can be adequately described by a Gaussian function. Hence, the glass fiber is included in the optical model by a second weighting function $W_{GF}(\varphi)$ of the form:

$$W_{GF}(\varphi) = \frac{1}{\sqrt{2\pi\sigma^2}} \exp\left(-\frac{1}{2} \cdot \left(\frac{3}{N}\sqrt{\frac{(2n-1)^2\varphi^2}{2}}\right)^2\right), \tag{1}$$

where $\varphi$ describes the angle of incidence, $\sigma$ the standard deviation of the Gaussian function, $n$ the control variable and $N$ the number of steps for discrete incidence angles used for describing the Gaussian weighting function (throughout this work, $N$ = 50). The 3$\sigma$ interval (99.73%) of the power of light is taken as scaling factor of this second weighting function. As the *NA* of the glass fiber is small (i.e. 0.07, which corresponds to an opening angle of 4°), it is neglected and the second weighting is achieved by a simple multiplication of the Gaussian function with the angular weighting function mentioned above[13]. In the latter, *N* is also in accordance to the number of incidence angles used for modelling of the *NA*. This mathematical operation corresponds to a convolution of both functions in frequency domain. Hence, the presented method is adaptable to largly arbitrary measurement setups by a modular describtion of numerical apertures of objective lenses by using the weighting factor $W_{NA}$[13] and glass fibers by including an additional weighting factor $W_{GF}$. Morover, the analytical model, which is based on the combination of the contrast-based and wavelength-shift based method, is widely capable of detecting arbitrary layer stacks on various technically relevant substrates. An explicit expression of the model is:

$$R(\lambda) = \sum_{n=1}^{N} R(\lambda, \varphi_n) \cdot W_{NA} \cdot W_{GF} = \sum_{n=1}^{N} R(\lambda, \varphi_n) \frac{2n-1}{N^2} \frac{1}{\sqrt{2\pi\sigma^2}} exp\left(-\frac{1}{2}\left(\frac{3}{N}\sqrt{\frac{(2n-1)^2 \varphi_n^2}{2}}\right)^2\right)$$

### 3.2. Verification of the suggested model

The optical model is verified by measurement of reflectance spectra of silicon dioxide layers with two different thicknesses by deploying four different objective lenses. In order to obtain more than one characteristic point (minimum/maximum) within the reflectance spectra, oxide layers with effective optical thicknesses above the maximum wavelength were chosen. Note, however, that the presented method is also applicable to an arbitrary (which can even include conductive materials) multilayer system featuring at least one extremum in the recorded wavelength range. As optical substrates two different layer stacks with silicon dioxide layers with nominal film thicknesses of 625 nm (thickness measured by SE: 626.5 ±0.9 nm) and 1.5 µm (thickness measured by SE: 1496.3 ± 2.1nm) on single-crystalline silicon are deployed. These layer stacks were fabricated under clean room conditions by wet-thermal oxidation of boron-doped silicon substrates with an acceptor concentration of $2 \cdot 10^{15}$ cm$^{-3}$, a crystal orientation of (100) and a diameter of 150 mm. The results of optical microspectroscopy are compared against the proposed optical model utilizing complex refractive indices of air, SiO$_2$ and silicon as depicted in **Figure 2.**

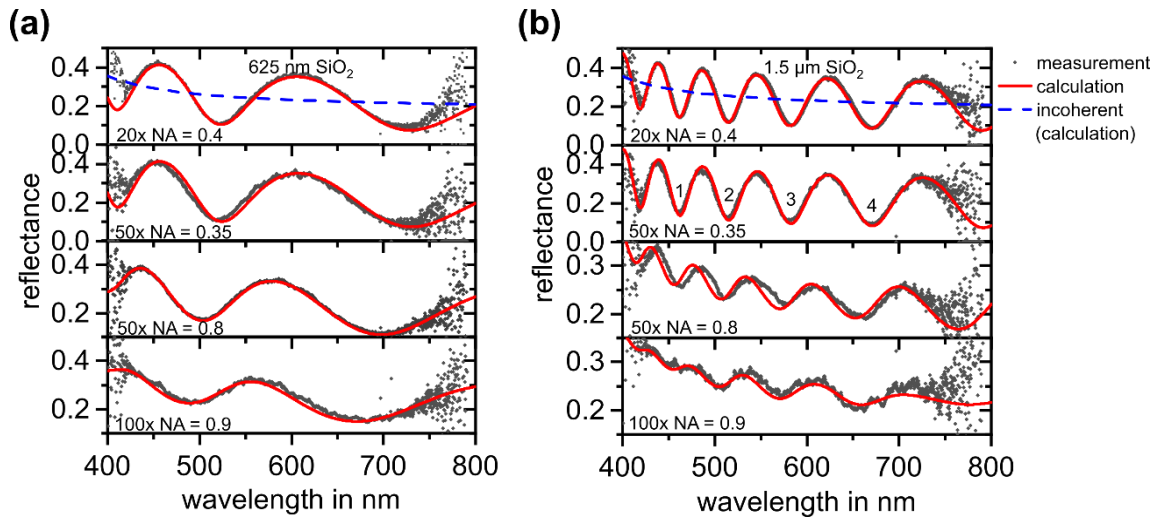

*Figure 2. Measured spectral reflectance of a silicon dioxide layer with a thickness of **(a)** 625 nm and **(b)** 1.5 µm on silicon substrates by utilizing four objective lenses with different magnification and NA (grey rhombi) and calculated spectra using the described optical model (red lines). Additionally, calculated incoherent reflectance spectra are given by blue, dashed lines.*

As can be seen in **Figure 2,** predicted spectra exhibit high conformance with measured spectra for both reference layers. Small deviations stem from low signal-to-noise-ratios (SNR) of the measured spectra at the detector as well as from small deviations in the film thickness of silicon dioxide of the specimen and the model. At this place it is explicitly mentioned that no fitting parameters are necessary as it is the case for the proposed models of Saigal et al.[13] (fit parameter: *NA*) and Katzen *et al.*[27] (fit parameter: polarization). In the present case, the introduced physical model is only supported by the results of layer thickness measurement using spectroscopic ellipsometry which is used for pre-characterization (standard process control) of thin film systems as well as by nominal device parameters of the instrument, i.e. the *NA* and the determined light distribution in the glass fiber.

### 3.3. Determination of the number of layers of Van-der-Waals heterostructures applying the proposed analytical model

In order to gain information about the precision of the analytical model the reflectance behavior of Van-der-Waals heterostructures is analyzed. For fabrication of samples, two different 2D materials, namely graphene and hBN, both purchased from ACS Material are deposited onto the substrates investigated above by PMMA-assisted transfer as described in literature[28–30]. The PMMA protection layer was removed by acetone exposure after deposition. **Figure 3(a)** shows an optical micrograph of Van-der-Waals heterostructures in different arrangements employing graphene and hBN. For characterization of spectral changes induced from the individual 2D materials, reflectance spectra were acquired from bare substrates, areas covered by graphene and hBN as well as from all overlapping areas (stacking order from substrate to top layer: $Si/SiO_2/Gr/hBN/Gr$). As hBN exhibits no visible contrast in this case, the position of the flake was determined by localizing PMMA residuals at the rim of the flake. In a more general case, spectral mapping is suggested as the strategy of choice for localizing barely visible materials. The measurement results were evaluated by determining spectral position and absolute reflectance of all occurring minima (and maxima, although these are not as well-defined as the minima and will, therefore, not be considered in the remainder). **Figures 3(b)-(f)** comprise results of this data evaluation scheme.

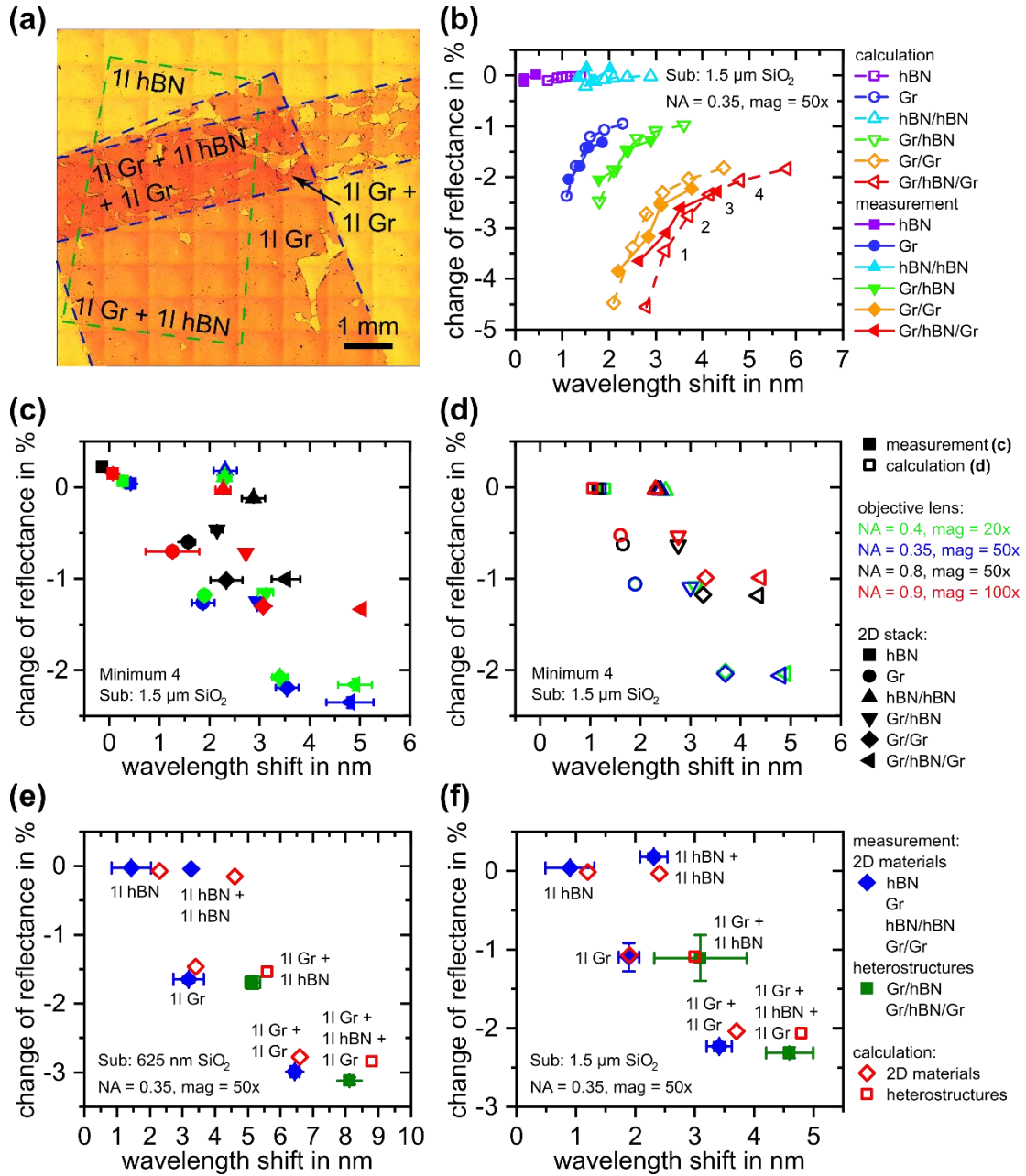

*Figure 3.(a)* large-field micrograph of Van-der-Waals heterostructures of graphene and hBN on 1,5 µm SiO$_2$ on silicon substrate (obtained by stitching and contrast enhancement), *(b)* comparison of calculated and measured deviation with respect to the bare substrate of minima of graphene-hBN heterostructures on 1.5 µm SiO$_2$ on silicon substrate for NA = 0.35 (see *Figure 2(b)* for the definition of the numbers of the minima), *(c)* measured and *(d)* calculated change of the fourth (cf. *Figure 2(b)*) reflectance minimum of graphene-hBN heterostructures on 1.5 µm SiO$_2$ on silicon substrate for different objective lenses (i.e. NAs) and *(e)* and *(f)* comparison of measured and calculated changes of the spectral position of reflectance minima of graphene-hBN heterostructures on 625 nm (first minimum) and 1.5 µm SiO$_2$ (fourth minimum) on silicon substrate.

**Figure 3(b)** displays a comparison of measurement results for the four distinct minima indicated in **Figure 2(b)** with calculated values for different Van-der-Waals heterostructures. Data were obtained through an objective lens with an *NA* of 0.35. The analytical calculations are based on the assumption of PMMA residuals of 3 nm in thickness for the case of graphene and of 2 nm for the case of hBN (with the residuals always on top of the 2D material). Those values are obtained by high resolution transmission electron microscopy (HRTEM) of a cross section prepared by focused ion beam (FIB) as well as atomic force microscopy (AFM) for the case of graphene[8] and the smaller PMMA thickness for hBN can be explained by the weaker interaction of PMMA with the polar material hBN featuring ionic bonds in contrast to the purely covalent graphene[31]. Utilizing these parameters calculated and measured values exhibit a high conformance indicating the supposed model to be precise enough for determining the number of layers of individual 2D materials in Van-der-Waals heterostructures featuring graphene and hBN. It is noticeable, that the wavelength shifts of extreme values located at shorter absolute wavelengths are smaller compared to wavelength shifts of minima at larger wavelengths. Vice versa, the relative change of reflectance is higher for minima at shorter absolute wavelengths. A remarkable observation is that graphene changes the spectral position as well as the absolute reflectance of the minima whereas hBN does only change the spectral position. This is due to the refractive index of hBN[21] being considerably lower than that of graphene[20]. Additionally, the extinction coefficient of hBN is negligible[21] which stands in contrast to the highly conductive material graphene that has a spectral extinction coefficient which strongly differs from zero within the investigated wavelength range[20]. This observation, furthermore, indicates that evidently the number of layers of hBN can only be determined by measuring the wavelength shift. An appropriate discrimination between both materials cannot be provided by only one parameter, the wavelength shift or the change in absolute reflectance. Consequently, for determination of the composition of Van-der-Waals heterostructures consisting of graphene and hBN only a combination of both evaluation strategies is capable as both parameters have to be taken into account.

Using the proposed evaluation scheme, the optical discrimination between graphene and hBN in heterostructures and the simultaneous determination of the number of their layers becomes accessible. Furthermore, as sufficient discriminability of hBN critically depends on the wavelength shift of extrema, the minimum accompanied with the largest wavelength shift (i.e. the fourth minimum) delivers the best sensitivity for thickness evaluation for further data evaluation. However, for other materials the choice of different extreme values might be advantageous. For example a trade-off between the magnitude of wavelength shift and change of reflectance could be achieved by utilizing extreme values where both sensitivities are sufficiently large because increasing one parameter results in a decrease of the other. Moreover, the choice of extrema at shorter wavelengths could be convenient for materials revealing no measurable wavelength shift utilizing the strong sensitivity of the reflectance. In the end, both parameters strongly depend on the complex refractive index of the respective material which has to be taken into account deliberately. This is especially important for 2D semiconductors[32], which exhibit oscillator frequencies within the measured spectral range due to direct band transitions[33]. Moreover, band transitions may change for some materials like black phosphorous[34] or molybdenum disulfide[35] depending on the exact number of layers.

**Figures 3(c)** and **3(d)** show the dependency of wavelength shift and deviation of reflectance on the *NA* of the utilized objective lens for measured and calculated spectra. For this particular minimum (i.e. the fourth minimum denoted in **Figure 2(b)**), the wavelength shift only slightly changes to lower values with increasing *NA* whereas the deviation of reflectance significantly decreases from 1.1% to 0.5% per layer of graphene by increasing the *NA* from 0.35 to 0.9. Measured values (**Figures 3(c)**) show high conformance to calculated values (**Figures 3(d)**) for *NA*s of 0.35 and 0.4. Even for a high *NA* of 0.9 both measured and calculated values are in good agreement even though lower SNRs caused by smaller amplitudes in reflectance spectra induce larger uncertainty. Only for an *NA* of 0.8 the error is quite large. This larger deviation might be caused by a lower baseline conformance which is also visible in **Figure 2(b)**. In addition, the conformance between calculation and measurement for NAs of 0.8 and 0.9 for the case of single layer hBN is quite low. Here, the precision of 1 nm is obviously not provided. This could be due to small ruptures in hBN at the measurement spots interfering the signal of hBN. Furthermore, the exact thickness of PMMA is not known which might deviate from assumed values locally and, thus, induces additional changes in reflectance spectra. In addition, slight changes in complex refractive index for different numbers of layers of 2D materials cannot be excluded. Nevertheless, a general trend, i.e. a wavelength shift for each hBN layer and an additional change in reflectance for each graphene layer is also visible for *NAs* of 0.8 and 0.9. Furthermore, as the error bars do not overlap, an unambiguous discrimination can be provided in each case.

The influence of the subjacent layer stack is investigated in **Figure 3(e)** and **3(f)**. Obviously, the amplitude of both parameters (i.e., wavelength shift and change of reflectance) is larger for the thinner oxide layer. The measured wavelength shift changes from 0.9 nm to 1.4 nm (calculated: 1.2 nm and 2.3 nm) for a single layer of hBN and from 1.9 nm to 3.2 nm (calculated: 1.9 nm and 3.4 nm) for monolayer graphene. For the latter, the deviation in reflectance decreases from -1.1% to -1.6% (calculated: -1.1% and -1.5%) for $SiO_2$ layers with a thickness of 1.5 µm and 625 nm, respectively. Although the deviation between measured and calculated values for the case of 625 nm of $SiO_2$ is remarkably larger, a sufficient discrimination between both 2D materials can still be provided. This is because hBN exhibits only a spectral shift of extreme values compared to the reflectance of the bare substrate whereas graphene is accompanied by an additional change in the absolute value of reflectance. A very high conformance between measured and calculated values is found for an optical substrate with 1.5 µm of $SiO_2$.

In order to verify the method for a broader range of applicable substrates, the justified model is extrapolated to a typical use case involving a rather thin subjacent $SiO_2$ layer of 90 nm (often used for TMDs like $MoS_2$, $WSe_2$ and $TaS_2$[36]). **Figure 4** shows the calculated spectra (**Fig. 4(a)**) and spectral changes (**Fig. 4(b)**) for 2D heterolayer stacks identical to the experiments depicted in **Figures 3(e)** and **(f)** above.

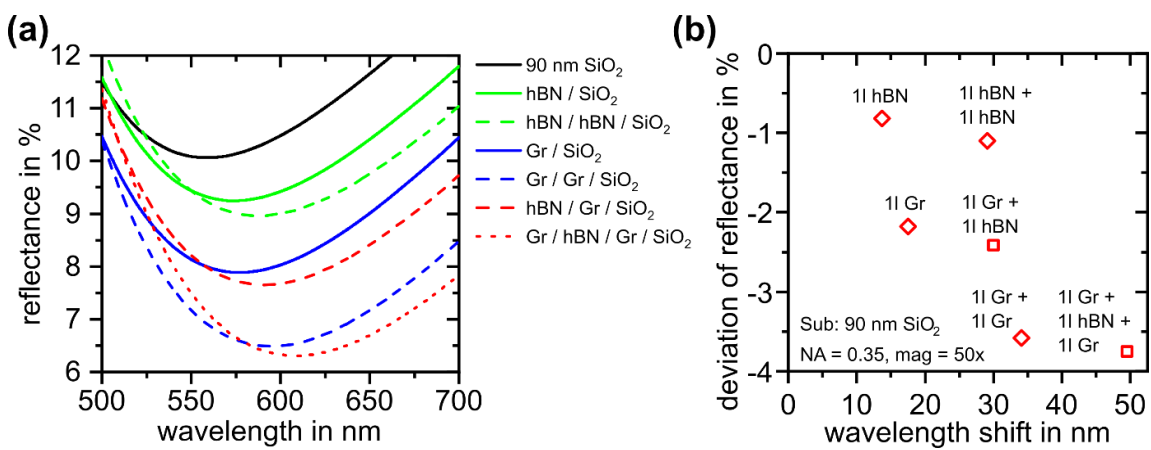

*Figure 4. (a)* Calculated reflectance spectra and *(b)* spectral changes of graphene/hBN homo- and heterostructures on a layer stack of 90 nm of silicon oxide on silicon substrate for NA = 0.35.

The calculations show that both wavelength shift and reflectivity changes are significantly pronounced with respect to thicker SiO$_2$ layers. While the absolute reflectance of the evaluated minimum is about 20% larger compared to the case of 625 nm SiO$_2$, the wavelength shift dramatically increases by a factor of more than five. Yet, evaluation of measurements performed on thin subjacent layers is less accurate as the minimum is rather wide and shallow. Noise might negatively interfere with the measurement data and might cause the evaluation of such data being more elaborate. Improved fitting algorithms for determining the wavelength of the minimum and long integration times for the measurements may deliver a work around for the inferior signal-to-noise ratios.

## 4. Conclusion

In this study, we introduce a combined data evaluation approach deploying the determination of both a wavelength shift as well as a change of reflectance of extreme values in reflectance spectra for discriminating different 2D materials in Van-der-Waals heterostructures. This advanced data evaluation strategy features an accurate physical model describing reflectance behavior of multilayer systems by a transfer-matrix method including *NA* and waveguide modelling. The validity of the proposed model is proven by conformance of reflectance spectra acquired via optical microspectroscopy and calculated spectra for reference samples with well-known refractive indices. Furthermore, it is shown that the precision of the analytical model is suitable for predicting wavelength shift and deviation in reflectance of extreme values enabling decisive discrimination of hBN and graphene in Van-der-Waals heterostructures. This decisive discriminability of both materials allows for conclusive allocation of respective measured and calculated spectral positions and absolute reflectances of extreme values within reflectance spectra. In the end, the proposed model overcomes restrictions of both individual methods evaluating just one parameter for layer stacks which do not exhibit a measurable contrast or wavelength shift.

**Data Availability**

The datasets generated during and/or analyzed during the current study are available from the corresponding author on reasonable request.

**Acknowledgement**

Financial support by the DFG via the Research Training Group GRK 1896 "In situ microscopy with electrons, X-rays and scanning probes" is gratefully acknowledged.

**Author Contributions**

A.H. wrote the main manuscript text, prepared all figures and performed specimen preparation, data acquisition and data evaluation as well as interpretation. A.H., C.M. and M.R. developed the optical model utilized for data evaluation. M.R., B.F., C.M. and A.H. coded the algorithms used for data evaluation. A.H., M.J and M.R. carefully revised the manuscript. All authors critically discussed the obtained data and reviewed the manuscript.

**Additional Information**

The authors declare no competing interests.